\documentclass[showpacs,pre]{revtex4}
\usepackage{amsmath}
\usepackage{graphicx}
\begin{document}
\title{Dynamical quorum sensing and clustering dynamics in a population of spatially distributed active rotators}
\author{Hidetsugu Sakaguchi and Satomi Maeyama}
\affiliation{Department of Applied Science for Electronics and Materials,
Interdisciplinary Graduate School of Engineering Sciences, Kyushu
University, Kasuga, Fukuoka 816-8580, Japan}
\begin{abstract}
A model of clustering dynamics is proposed for a population of spatially distributed active rotators.  A transition from excitable to oscillatory dynamics is induced by the increase of the local density of active rotators. It is interpreted as dynamical quorum sensing. In the oscillation regime, phase waves propagate without decay, which generates an effectively long-range interaction in the clustering dynamics. The clustering process becomes facilitated and only one dominant cluster appears rapidly as a result of the dynamical quorum sensing. An exact localized solution is found to a simplified model equation, and the competitive dynamics between two localized states is studied numerically.   
\end{abstract}
\maketitle
Synchronization in populations of coupled oscillators have been extensively studied \cite{rf:1} since the pioneering works by Winfree \cite{rf:2} and Kuramoto~\cite{rf:3}.
In spatially extended systems such as the BZ reaction, target patterns and spiral patterns appear as a result of synchronization~\cite{rf:4}. 
On the other hand, populations of unicellular organisms sometimes undergo drastic transitions when the cell density increases. For example, some kinds of bacteria change to produce toxic substances when the cell density is beyond a threshold.  
 The phenomenon is called quorum sensing in the meaning that the transition occurs by sensing the density of the same kind of organisms. Dynamical quorum sensing is used for a phenomenon that populations of active elements undergo some dynamical transitions with increasing the density. Dynamical quorum sensing was studied for synthetic multi-cellular clock \cite{rf:5}, cell-density-dependent glycolytic oscillation in yeast \cite{rf:6}, and chemical oscillators \cite{rf:7}. Gregor et al. studied collective behaviors of social amoebae: Dictyostelium discoideum, and found a transition from an excitable state to an oscillatory one with increasing the cell density\cite{rf:8}. The oscillation of extra cellular cAMP is observed in the experiment. The social amoebae begin to aggregate when the period of the oscillation becomes shorter. Some local aggregates are created initially, but a competition among local aggregates occurs via cell-cell signaling. The cell-cell signaling is caused by the waves of cAMP.  Finally, only one dominant aggregate appears.  

There is a mathematical model for the oscillation of cAMP by Martiel and Goldbeter \cite{rf:9}. However, Gregor et al. used a simpler phase equation to explain the transition from an excitable state to an oscillatory state. 
  We use a similar type of phase equation which can express a transition from excitatory dynamics to oscillatory one:
\begin{equation}
\frac{d\phi}{dt}=\omega-b\sin\phi.
\end{equation}
When $\omega<b$, there is a stable stationary solution $\phi=\sin^{-1}(\omega/b)$. The stationary state can be regarded as excitable because it is easily excited by a small external input when $\omega$ is close to $b$. When $\omega>b$, the phase increases monotonically, which represents the limit-cycle oscillation. 
Gregor et al. studied a globally coupled system of the form:
\begin{equation}
\frac{d\phi_i}{dt}=\omega\left (1-\frac{B\sin\phi_i}{K+C\sum_{j=1}^N\{(-A_m+A_b)\sin\phi_j+A_m+A_b\}/N}\right ).
\end{equation}
In this paper, we consider a spatially-extended oscillatory medium in one dimension. We use a phase equation of the form
\begin{equation}
\frac{\partial\phi}{\partial t}=\omega-b\sin\phi+\nu\frac{\partial^2\phi}{\partial x^2}+g\left (\frac{\partial \phi}{\partial x}\right )^2,
\end{equation}
because this type of phase equation is more general, although Eq.~(2) or its generalization might be better for the quantitative argument. The third term expresses the phase diffusion, and the last nonlinear term $g(\partial \phi/\partial x)^2$ plays an important role in the wave propagation such as target and spiral patterns especially in oscillatory systems. The derivation of the diffusion term and the nonlinear term for general oscillatory systems was discussed in Ref.[3].   
We use a term "active rotator" in this paper which exhibits excitable or oscillatory dynamics and moves actively by sensing its density.
We assume that $\omega$ in Eq.~(3) is a function of $n$ as $\omega=\alpha n/(1+\gamma n)$, taking into account an experimental fact that the frequency increases with the cell density of the social amoebae and is saturated when the density is sufficiently large.

In this paper, we would like to study interaction between the dynamical quorum sensing and the aggregation dynamics. The Keller-Segel model is a model of the aggregation dynamics induced by chemotaxis~\cite{rf:10,rf:11}.  In the chemotaxis, the organisms move, sensing the concentration of chemical substances. 
In a variant of the  Keller-Segel model, the density $n(x,t)$ obeys 
\begin{equation}
\frac{\partial n}{\partial t}=D\frac{\partial^2n}{\partial x^2}-\frac{\partial}{\partial x}\left (\frac{\partial \chi(v)}{\partial x}n\right ),
\end{equation}
where $D$ is the diffusion constant for $n$ and $\chi(v)$ represents a sensitivity function for the concentration $v$ of the chemical substance. The organism moves toward the region of high $\chi(v)$. Oscillatory dynamics is not explicitly taken into consideration in the Keller-Segel model. Van Oss et al. studied a complicated model combining the Martiel-Goldbeter model of the cAMP oscillation with a certain aggregation dynamics, and succeeded in reproducing the motion toward the center of a spiral pattern, i.e., the source of the cAMP waves.

In this paper, we use a variant of Eq.~(4) for the aggregation dynamics.  In our model,  $\chi(v)$ in Eq.~(4) is replaced by $\phi$. It is because the cells move in the opposite direction to the propagation of cAMP waves. As a result they tend to aggregate toward the source of cAMP waves. The waves are sent out from a pacemaker region where the phase $\phi$ is in advance. 

The purpose of this paper is to show that the aggregation dynamics changes qualitatively by the dynamical quorum sensing. 
Our model equations are expressed as 
\begin{eqnarray}  
\frac{\partial\phi}{\partial t}&=&\frac{\alpha n}{1+\gamma n}-b\sin\phi+\nu\frac{\partial^2\phi}{\partial x^2}+g\left (\frac{\partial \phi}{\partial x}\right )^2,\\
\frac{\partial n}{\partial t}&=&D\frac{\partial^2n}{\partial x^2}-\frac{\partial}{\partial x}\left (\frac{\partial \phi}{\partial x}n\right ). \label{model}
\end{eqnarray} 
We study the coupled model equations numerically and theoretically. 
For numerical simulations, periodic boundary conditions $\phi(0)=\phi(L)$ and $n(0)=n(L)$ are imposed. The total number of active rotators $N=\int_0^L n(x,t)dx$ is conserved in the time evolution of Eq.~(\ref{model}).  
We will study the model equations (5) and (6) to understand the interaction between the dynamical quorum sensing and the aggregation dynamics qualitatively in this paper.  However, it is noted that the model equations (5) and (6) have a limit of application for quantitative argument: the amplitude of the oscillation is assumed not to change vastly in the phase description, but the oscillatory dynamics of cAMP waves may change qualitatively when the density $n(x,t)$ becomes quite low, and the phase description might become worse.  

There is a uniform and stationary solution $\phi(x,t)=\phi_0$ and $n(x,t)=n_0=N/L$, which satisfies 
\begin{equation}
\frac{\alpha n_0}{1+\gamma n_0}=b\sin\phi_0,
\end{equation}
if $b>\alpha n_0/(1+\gamma n_0)$. Small perturbations $\delta \phi=\phi-\phi_0$ and $\delta n=n(x,t)-n_0$ obey linear equations:
\begin{eqnarray}  
\frac{\partial\delta\phi}{\partial t}&=&\frac{\alpha \delta n}{(1+\gamma n_0)^2}-b\cos\phi_0\delta\phi+\nu\frac{\partial^2\delta\phi}{\partial x^2},\nonumber\\
\frac{\partial \delta n}{\partial t}&=&D\frac{\partial^2\delta n}{\partial x^2}-\frac{\partial}{\partial x}\left (\frac{\partial \delta \phi}{\partial x}n_0\right ).
\end{eqnarray} 
By the Fourier transform, the Fourier amplitudes $\delta \phi_k$ and $\delta n_k$ with wavenumber $k$ obey
\begin{eqnarray}  
\frac{\partial\delta\phi_k}{\partial t}&=&\frac{\alpha \delta n_k}{(1+\gamma n_0)^2}-b\cos\phi_0\delta\phi_k-\nu k^2\delta\phi_k,\nonumber\\
\frac{\partial \delta n_k}{\partial t}&=&-Dk^2\delta n_k+n_0k^2\phi_k.
\end{eqnarray} 
The eigenvalue $\lambda_k$ is evaluated at
\begin{equation}
\lambda_k=\frac{-(Dk^2+\nu k^2+b\cos\phi_0)+\sqrt{(Dk^2-\nu k^2-b\cos\phi_0)^2+4\alpha n_0k^2/(1+\gamma n_0)^2}}{2}\sim a_1k^2+a_2k^4,
\end{equation}
where 
\[a_1=-D+\frac{\alpha n_0}{b\cos\phi_0(1+\gamma n_0)^2},\;a_2=-\frac{\alpha^2n_0^2}{(1+\gamma n_0)^4b^3\cos^3\phi_0}+\frac{(D-\nu)\alpha n_0}{(1+\gamma n_0)^2b^2\cos^2\phi_0}\]
for sufficiently small $k$. The uniform state becomes unstable for $D<\alpha n_0/\{b\cos\phi_0(1+\gamma n_0)^2\}$,  and active rotators  begin to aggregate locally.  

\begin{figure}[t]
\begin{center}
\includegraphics[height=4.5cm]{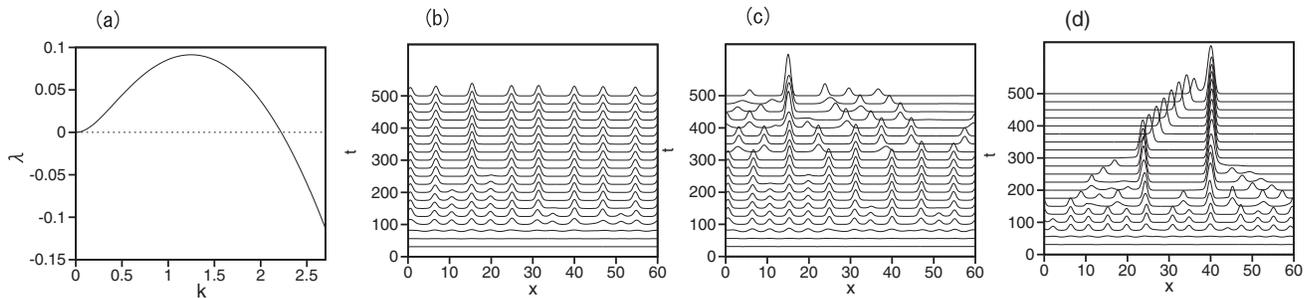}
\end{center}
\caption{(a) Eigenvalue $\lambda(k)$ for $\alpha=0.15$. Time evolutions of $n(x,t)$ at (b) $\alpha=0.15$, (c) 0.16, and (d) 0.18  for $\gamma=0.5, \nu=0.1, g=1, b=0.2, D=0.1$ and $L=60$.}
\label{f1}
\end{figure}
Figure 1(a) shows the eigenvalue $\lambda(k)$ by Eq.~(10) at $\alpha=0.15$. The other parameters are fixed to be $\gamma=0.5,\,\nu=0.1,\, g=1, \,b=0.2, \,D=0.1,\,n_0=1$ and $L=60$. The eigenvalue takes a maximum near $k=k_m=1.25$. 
Figures 1(b)-(d) show time evolutions of $n(x,t)$ at (b) $\alpha=0.15$ and (c) $\alpha=0.16$  and (d) $\alpha=0.18$. The initial conditions are $\phi(x,0)=0$ and $n(x,0)=1+r(x)$ where $r$ is a random number between -0.02 and 0.02. The uniform state is unstable, a wavy pattern appears. The number of peaks near $t=100$ at $\alpha=0.15$ is 12, and therefore the average wavelength is evaluated at $60/12=5$, which is close to $2\pi/k_m=5.02$ for $\alpha=0.15$ shown in Fig.~1(a). 
The linear stability analysis explains the length scale of initial wavy patterns. The wavy pattern develops into localized clusters of active rotators.  
The clusters are mutually merged in $100<t<200$, and the cluster number decreases during the time. 
The localized clusters become almost stationary at $\alpha=0.15$. 
However, a rapid merging process among clusters occurs at $t>300$ for $\alpha=0.16$. 
We have shown only for $t<500$ in Fig.~1(c), however, the merging process continues and one cluster appears finally. 
The merging process occurs more strongly at $\alpha=0.18$, and only one cluster dominates at $t=500$.   
Figure 2(a) shows time evolutions of $\phi(x_0)$ at $x_0=15$ for $\alpha=0.15,\,0.16$ and  0.18. 
The phase $\phi(x_0)$  remains to be below $\pi/2$ for a fairly long initial time, but it begins to increase abruptly at $t\sim 300$ for $\alpha=0.16$ and at $t\sim 140$ for $\alpha=0.18$, which implies that a transition from an excitable state into an oscillatory state has set in at the times. Stepwise time evolutions are seen for $\alpha=0.16$ and 0.18 in the oscillatory state. It is because the time derivative $\partial \phi/\partial t$ becomes small when $\phi$ goes through $\pi/2+2\pi n$. Figure 2(b) shows the time evolution of the profile $\phi(x,t)$ at $\alpha=0.16$. The overlapped profiles below $\phi=\pi/2$ correspond to $\phi(x)$'s for $t<300$.  The phase patterns grow abruptly at $t\sim 300$ by the transition.  The phase profile has a peak around $x\sim x_0=15$, which implies that phase waves are sent out from $x\sim x_0$.  Because the active rotators move to the center of the phase waves in our model, the localized clusters move toward the center and they are absorbed into a dominant cluster near $x=x_0$ at $\alpha=0.16$ as shown in Fig.~1(c). At $\alpha=0.15$,  the transition to the oscillatory state does not occur and therefore the rapid merging of localized clusters is not observed. 
We can interpret the transition from the excitable state to the oscillatory state as a kind of dynamical quorum sensing in that the transition occurs when the local density increases and goes beyond a threshold. The clustering process changes qualitatively by the dynamical quorum sensing. 
In the excitatory regime, the clustering proceeds locally, however, a rapid clustering into one dominant cluster does not occur. This type of local clustering is similar to that in the Keller-Segel models~\cite{rf:13}, in that the oscillatory dynamics is not involved. 
In the oscillation regime, localized clusters are attracted to the center of phase waves and are absorbed into the dominant cluster even if the localized clusters are far distant from the dominant cluster.  It is because the phase waves propagate far away without decay in the oscillation regime.  
\begin{figure}[t]
\begin{center}
\includegraphics[height=4.cm]{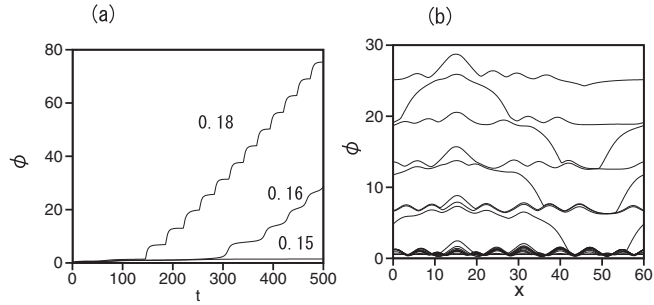}
\end{center}
\caption{(a) Time evolutions of $\phi(x,t)$ at $x=x_0=15$  for $\alpha=0.15,0.16$ and 0.18. (b) Time evolution of $\phi(x,t)$ at $\alpha=0.16$. The other parameters are $\gamma=0.5,\,\nu=0.1,\, g=1, \,b=0.2, \,D=0.1$ and $L=60$.
}
\label{f2}
\end{figure}

The transition from the excitable state to the oscillatory state can be studied more easily for a one-hump state. In the excitatory regime, there is a localized stationary solution, which has a form: 
\begin{equation}
\phi(x)=\phi_s-ax^2,\; n(x)=n_se^{-cx^2}, \label{st}
\end{equation}
where $\phi_s,a, n_s$ and $c$ are parameters to characterize the form of the one-hump solution.  Substitution of Eq.~(\ref{st}) into Eq.~(5) yields
\begin{equation}
\frac{\alpha n_s}{1+\gamma n_s}-b\sin\phi_s-2\nu a=0,\;\frac{\alpha n_sc}{(1+\gamma n_s)^2}-ba\cos\phi_s-4ga^2=0,  \label{st2}
\end{equation}
by comparing the coefficients of the first and second terms in the Taylor expansion by $x$ around $x=0$. Furthermore, $Dc=a$ is satisfied from the relation $D\partial n/\partial x=(\partial \phi/\partial x)n$ in the stationary state. The total number is expressed as $N=\int_{-\infty}^{\infty}n(x)dx=n_s\sqrt{\pi/c}$. Then, the parameters $a$ and $c$ are expressed as $a=\pi n_s^2D/N^2$ and $c=\pi n_s^2/N^2$. Substitution of these relations into Eq.~(\ref{st2}) yields coupled equations for $\phi_s$ amd $n_s$. We can solve the coupled equations numerically. The solution is $n_s=4.92$ and $c=1.55$ for $\alpha=0.16,\gamma=1,g=1,\nu=0.1,D=0.1$ and $N=7$. Figure 3(a) shows $n(x,t)$ (solid curve) in the stationary state by direct numerical simulation of Eqs.~(5) and (6) and $n_se^{-cx^2}$ (dashed curve) with $n_s=4.92$ and $c=1.55$ for the same parameters: $\alpha=0.16,\gamma=1,g=1,\nu=0.1,D=0.1$ and $N=7$. Fairly good agreement is seen. Figure 3(b) shows a relation of $\phi_s$ and $N$ for $\alpha=0.16,\gamma=0.5,g=1,\nu=0.1$ and $D=0.1$ by the direct numerical simulation (rhombi) and by numerical results (solid curve) using Eq.~(\ref{st2}). The two values of $\phi_s(N)$ are in good agreement. The stationary solution $\phi_s$ increases with $N$ and reaches $\pi/2$, and then the stationary solution disappears when $N$ is greater than the critical value. 
It implies the transition to the oscillatory state. 
  
\begin{figure}[t]
\begin{center}
\includegraphics[height=4.cm]{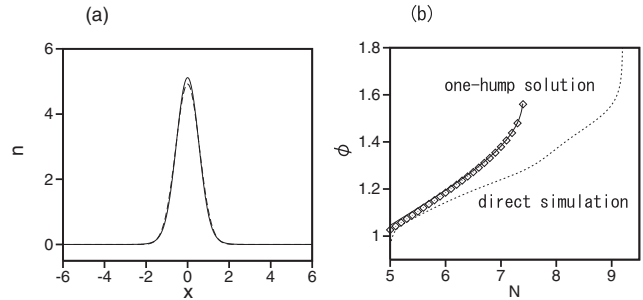}
\end{center}
\caption{(a) Stationary solutions $n(x,t)$ (solid curve)  by direct numerical simulation of Eqs.~(5) and (6), and $n_se^{-cx^2}$ (dashed curve) for $\alpha=0.16,\gamma=1,g=1,\nu=0.1,D=0.1$ and $N=7$. (b) Relations of the peak value $\phi_s$ and $N$. Solid curve shows numerical results by  Eq.(\ref{st2}), and  rhombi denote  direct numerical results for $\alpha=0.16,\gamma=0.5,g=1,\nu=0.1$ and $D=0.1$. The dashed curve denotes a relation of the maximum value $\phi_m$ of $\phi(x,t)$ in $13<x<17$ and $N_l=\int_{13}^{17}n(x,t)dx$ in the time evolution of direct numerical simulation shown in Fig.~1(c).} 
\label{f3}
\end{figure}
In the time evolution shown in Fig.~1(c), there is a localized structure of $n(x,t)$ around $x=x_0\sim 15$. The local density $n(x,t)$ increases in time there. We have calculated the number of active rotators in the localized structure as $N_l=\int_{x_1}^{x_2}n(x,t)dx$ where $x_1=13$ and $x_2=17$ and the maximum value $\phi_m$ of $\phi(x,t)$ in $13<x<17$.  $N_l$ increases monotonically, and $\phi_m$ also increases with $N_l$. The two values reach $N_l=9.05$ and $\phi_m=1.57$ at $t=267$, and then the transition into an oscillatory state occurs.  A relation of $N_l$ and the phase $\phi_m$ in the time evolution is shown in Fig.~3(b) by a dashed curve. The trajectory of $(N_l,\phi_m)$ follows the curve of the stationary solution for $N<6$, which implies the localized structure is close to the stationary solution, however, the trajectory deviates from that of one-hump solutions for $N>6$, and $N_l$ increases to 9.05 and then a transition to the oscillatory state occurs.  The deviation might be due to the non-stationarity or the rapid increase of $N_l$. 
In any case, the critical value of $\phi_m$ is around $\phi_m\sim \pi/2=1.57$ even for this non-stationary localized structure.    
\begin{figure}[tbp]
\begin{center}
\includegraphics[height=4cm]{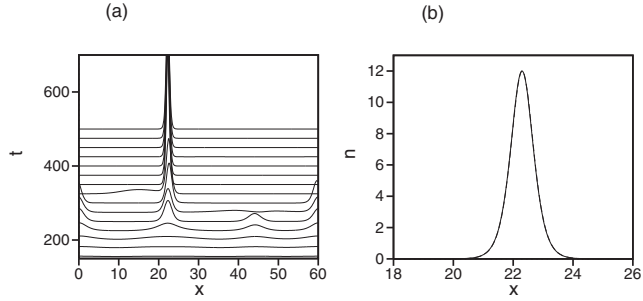}
\end{center}
\caption{(a) Time evolution of $n(x,t)$ for $b=0, \gamma=0, \alpha=0.5, g=1$ and $N=12$. (b) Stationary solutions obtained by the direct numerical simulation (solid curve) and theoretical one (dashed curve). The difference is invisible. }
\label{f4}
\end{figure}

In the oscillation regime, many localized structures compete with each other, and only one localized structure survives finally. The dynamics in the oscillation regime can be studied using a simpler system of $b=0$ and $\gamma=0$. Figure 4(a) shows the time evolution of $n(x,t)$ for $\alpha=0.5$, $\nu=0.5,g=1, D=0.5$ and $N=12$. The initial condition is $\phi(x,0)=0$ and $n(x,t)=0.2+r(x)$ where $r$ is a random number between -0.001 and 0.001. 
Three localized structures are created initially, and only one cluster survives by the competition among the three localized structures. Figure 4(b) shows the stationary solution localized around $x=22.3$. The stationary solution can be explicitly solved in case of $b=0$ and $\gamma=0$ by the ansatz $n(x)=A/\cosh^2(kx)$ and $\phi(x)=D\ln n+\omega t$. 
Substitution of the ansatz into Eqs.~(5) and (6) determines the parameter $A,k$ and $\omega$ as
\begin{equation}
k=\frac{N\alpha}{4\nu D+8gD^2},\;A=\frac{N^2\alpha}{8\nu D+16gD^2},\;\omega=\frac{gN^2\alpha^2D^2}{(2\nu D+4gD)^2}.
\end{equation}
This solution is an exact solution of the nonlinear equations (5) and (6). 
The parameters are determined as $k=2$ and $A=12$ for $\alpha=0.5,g=1$ and $N=12$. The dashed curve in Fig.~4(b) is the theoretical one: $n(x)=12/\cosh^2\{2(x-22.3)^2\}$. The difference is invisible. 
 
\begin{figure}[tbp]
\begin{center}
\includegraphics[height=4.cm]{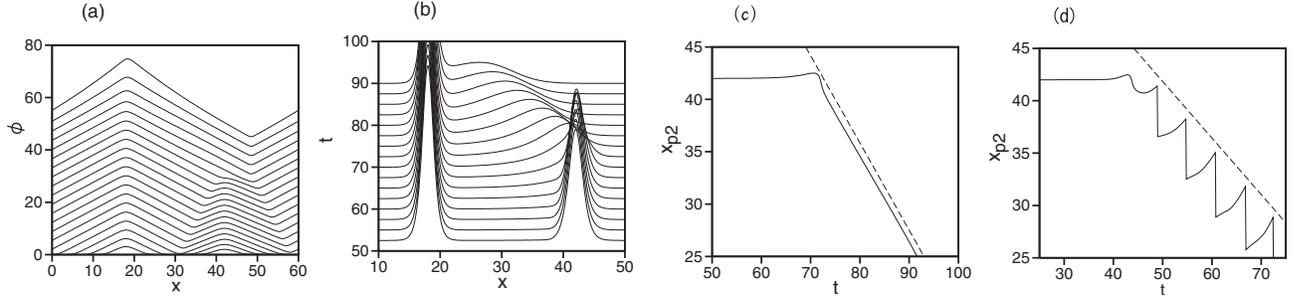}
\end{center}
\caption{(a) Time evolution of $\phi(x,t)$. The initial conditions are $n(x,0)=A_1/\cosh^2\{k_1(x-x_1)\}+A_2/\cosh^2\{k_2(x-x_2)\}$ where $A_1=2.0833,k_1=0.8333, x_1=18$ and $A_2=1.333,k_2=0.666$, and $x_2=42$  (b) Time evolution of $n(x,t)$ in the region of $10<x<50$ for $50<t<90$. (c) Time evolution of the peak position $x_{p2}$ of the second localized state. The dashed line is a line of velocity $-0.833$. (d) Time evolution of the peak position $x_{p2}$ of the second localized state for $b=0.25$ and $\gamma=0$. The dashed line is a line of velocity $-0.45$.}
\label{f5}
\end{figure}
To understand the competitive dynamics between two clusters, we have performed a numerical simulation from an initial conditions: $\phi(x,0)=0$ and $n(x,0)=A_1/\cosh^2k_1(x-x_1)+A_2/\cosh^2k_2(x-x_2)$ where $A_1=2.0833,k_1=0.8333, x_1=18$ and $A_2=1.333,k_2=0.666, x_2=42$. The parameters $A_1,k_1,A_2$ and $k_2$ are theoretical values by Eq.~(12) respectively for $N_1=5$ and  $N_2=4$ at $\alpha=0.5,\nu=D=0.5$ and $g=1$. 
Figure 5(a) shows time evolution of $\phi(x,t)$. Initially, phase waves are sent out from the two center $x_1=18$ and $x_2=42$. 
The phase waves from the two centers are expressed as
\begin{equation}
\phi_i=D\ln n_i(x)+\omega_1t=D\ln\{A_i/\cosh^2(k_ix)\}+\omega_i t\sim \pm 2Dk_ix+D\ln (4A_i)+\omega_i t,\;\; {\rm for}\; x\rightarrow \mp \infty,
\end{equation}
for $i=1$ and 2. 
The two phase waves are approximated at $\phi_1\sim -2Dk_1(x-x_1)+\omega_1 t$ and $\phi_2\sim 2Dk_2(x-x_2)+\omega_2 t$ in the region of $x_1<x<x_2$. 
The region $x_1<x<x_2$ is separated into two regions by a boundary or a shock where the two phase waves collide with each other. The two localized structures are fairly stable when the boundary is far from the two centers. 
However, the frequency $\omega_1$ of the left waves is faster and therefore the boundary (the shock) between the two regions moves in the right direction.  The boundary point is evaluated as $x=(\omega_1-\omega_2)t/\{2D(k_1+k_2)\}+(k_2x_2+k_1x_1)/(k_1+k_2)$ from the relation $\phi_1=\phi_2$, and therefore the velocity of the boundary is evaluated as $v=(\omega_1-\omega_2)/\{2D(k_1+k_2)\}$. When the boundary reaches $x_2$, the localized structure of the right cluster begins to be destroyed. Figure 5(b) shows the time evolution of $n(x,t)$ in the region of $10<x<50$ for $50<t<90$. The peak of the second localized structure moves to the left, the peak height decreases, and finally the right cluster is swallowed by the left dominant cluster.  
Figure 5(c) shows the peak position of the second (right) localized structure. 
The peak position is almost stationary until $t=70$, but moves to the left for $t>72$. The dashed line shows $x\sim -0.833 t$, that is, the velocity of the second peak is approximately $-2Dk_1\sim -0.833$. It is because the time evolution of $n(x,t)$ around the second peak is approximated by
\begin{equation}
\frac{\partial n}{\partial t}=D\frac{\partial^2 n}{\partial x^2}-\frac{\partial\phi}{\partial x}\frac{\partial n}{\partial x}-\frac{\partial^2\phi}{\partial x^2}n\sim  D\frac{\partial^2n}{\partial x^2}+2Dk_1\frac{\partial n}{\partial x},
\end{equation}
owing to $\partial \phi/\partial x\sim -2D k_1$. Here, the phase gradient $\partial\phi/\partial x$ is approximated at the phase gradient $-2Dk_1$ of the phase wave sent out from the left center as is seen from Fig.~5(a). 
 The last term $2Dk_1(\partial n/\partial x)$ induces the motion of the second peak with velocity $-2Dk_1$, and the first term makes the localized structure diffuse.  
Finally, a localized state appears around $x=x_1$ with total number $N=4+5=9$. 
Phase waves with a constant phase gradient propagate to the regions far away from the center, which is an origin of the effective long-range interaction in the oscillation regime. 
The effective long-range interaction in the oscillation regime facilitates the clustering process, and the clusters which are initially distant from one another are merged into one cluster rapidly. 

The approximation of the average velocity at $-2D k_1$ of the second peak is valid even for nonzero $b$ if $b$ is small, however, this approximation becomes worse near the transition to the excitable system. Figure 5(d) shows the peak (maximum) position of the second localized structure at $b=0.25$ for $\gamma=0, N_1=5,N_2=4, \alpha=0.5,\nu=D=0.5$ and $g=1$. The second localized structure is destroyed, a small two-peak structure appears, the two-peak structure exhibits oscillation, and moves toward the first localized structure on the average. The average velocity $v_2$ of the two-peak structure is 0.54, however, the average phase gradient $k_1$ is about 0.9, and the equality $v_2=-2Dk_1$ is not satisfied at this parameter. The discontinuous time evolution in Fig.~5(d) is due to the two-peak structure. 

In summary, we have proposed a model of clustering dynamics of spatially-distributed active rotators.  A transition from an excitable state to an oscillatory state is induced by the increase of local density, which is interpreted as the dynamical quorum sensing. The transition occurs when the peak value of $\phi(x,t)$  goes over $\pi/2$. The clustering dynamics changes qualitatively by the dynamical quorum sensing. That is, the clustering process becomes rapid, and only one cluster survives after the competition among localized clusters.
 We have found an exact localized solution for a simpler system of $\gamma=b=0$, and studied the competitive dynamics between two localized states. 
In this paper, we have used a rather simplified model to understand the dynamics qualitatively. We would like to study a more quantitative model equation in the future, using a modified model of Eq.~(2) and taking into account experimental results. It is another problem to extend the one-dimensional model equations (5) and (6) to a two-dimensional system.


\begin{thebibliography}{99}
\bibitem{rf:1} A.~Pikovsky, M.~Rosenblum and J.~Kurth, {\it Synchronization} (Cambrdge University Press, Cambridge, 2001).
\bibitem{rf:2} A.~T.~Winfree, J. Theor. Biol. {\bf 16}, 15 (1967).
\bibitem{rf:3} Y.~Kuramoto, {\it Chemical Oscillations, Waves and Turbulence} (Springer-Verlag, New York, 1984).
\bibitem{rf:4} A.~T.~Winfree, {\it The Geometry of Biological Time} (Springer-Verlag, New York, 1980).
\bibitem{rf:5} J.~Garcia-Ojalvo, M.~B.~Elowitz, and S.~H.~Strogatz, PNAS {\bf 101}, 10955 (2004).
\bibitem{rf:6} S.~De Monte, F.~d'Ovidio, S.~Dane, and P.~G.~Serensen, PNAS {\bf 104}, 18377 (2007).
\bibitem{rf:7} A.~F.~Taylor, M.~R.~Tinsley,F.~Wang, Z.~Huang, and K.~Showalter, Sicence {\bf 323}, 614 (2009).
\bibitem{rf:8} T.~Gregor, K.~Fujimoto, N.~Masaki, and S.~Sawai, Science {\bf 328}, 1021 (2010).
\bibitem{rf:9} J.~L.~Martiel and A.~Goldbeter, Biophys. J. {\bf 52}, 807 (1987).
\bibitem{rf:10} E.~F.~Keller and L.~A.~Segel, J. Theor. Biol. {\bf 26}, 399 (1970).
\bibitem{rf:11} V.~Nanjundiah, J. Theor. Biol. {\bf 42},63 (1973).
\bibitem{rf:12} C.~Van Oss, A.~V.~Panfilov, P.~Hogeweg, F.~Siegert, and C.~J.~Weijer, J. Theor. Biol. {\bf 181}, 203 (1996).
\bibitem{rf:13} T.~Hillen and K.~J.~Painter, J. Math. Biol. {\bf 58}, 183 (2009).  
\end{thebibliography}
\end{document}